\documentstyle [12pt]{article}
\title{{\bf TC corrections to the single-top-quark production at 
the Fermilab Tevatron  }}
\author{Gongru Lu, Yigang Cao, Jinshu Huang, Junde Zhang \\
{\small  Physics Department, Henan Normal University, Xinxiang, Henan, 
453002, P. R. China\thanks{E-mail: Lugr@sun.ihep.ac.cn} } \\ 
Zhenjun Xiao\thanks{E-mail: xiaozj@v2.rl.ac.uk} \\ 
{\small Rutherford Appleton Laboratory, 
Chilton, DIDCOT, OX11 0QX, U.K. \thanks{Current mailing address}}\\
{\small  Physics Department, Henan Normal University, Xinxiang, Henan,
453002, P. R. China } \\ }
\date{}

\begin{document}
\maketitle
\begin{picture}(0,0)(0,0)
\put(340,265){{\large HNU-TH/97-01}}
\put(340,250){{\large HEP-PH/9701xxx}}
\end{picture}
\begin{abstract}        
 We calculate one-loop corrections to the single-top-quark production via 
$q\overline{q}' \rightarrow t\overline b$ at the Fermilab Tevatron from 
the Pseudo-Goldstone bosons ( PGBs ) in the framework of 
one generation technicolor model. 
The maximum correction to the total cross section 
for the single-top-quark production is found to reach -2.4\% relative to 
the tree-level cross section, which may  be observable at a high-luminosity 
Tevatron. 

\end{abstract}

\vspace{0.5cm}

PACS numbers: 12.15.LK,12.60.Nz,13.30.Eg

%

\newpage
\begin{flushleft}    
\section*{I. Introduction}
\end{flushleft}

Recently, the top quark was discovered at the Tevatron with the mass $m_t=
174.4 \pm 8.3 GeV$ [1], which is of the order of the electroweak 
symmetry breaking ( EWSB ) scale $\mu=(\sqrt{2} G_F )^{-1/2}=246 GeV$. This 
means that the top quark couples rather strongly to the EWSB sector so that 
the effects from new physics would be more apparent in the  processes with 
the top quark than with other light quarks [2]. Experimentally, it is 
possible to measure all of the production and decay form factors of the top 
quark at the level of a few percent separately [3]. Therefore, theoretical 
calculations of the radiative corrections to the production and decay of the 
top quark is of much interest. In this paper, we will address the production 
of top quark.

At the Tevatron top quark are produced primarily via two independent 
mechanisms: The dominant production mechanism is the QCD pair production 
process $q\overline{q}\rightarrow t\overline{t}$ [4]. W-gluon fusion process 
$g+W\rightarrow t\overline b$ [5, 6] and Drell-Yan type process 
$q\overline{q}'\rightarrow t\overline{b}$ [7] are important too. 
These latter processes produce a single top quark, rather than a 
$t\overline {t}$ pair. Both involve 
the weak interaction, so they are suppressed relative to the strong 
production of $t\overline{t}$; however, this suppression is partially 
compensated by the presence of only one heavy particle ( such as the top 
quark ) in the final state. Both processes probe the charged-current 
weak interaction of the top quark. And it was shown in Ref.[8] that the 
signal for single top production in $q\overline{q}'\rightarrow t\overline{b}$ 
via a timelike W boson $q^2 > (m_t+m_b)^2$ is potentially observable at the 
Tevatron. The signal for this process is unobservable at LHC due to the large 
background from $t\overline{t}$ production and single top production via 
W-gluon fusion [6]. Compared to the single top production via W-gluon fusion 
the process $q\overline{q}'\rightarrow t\overline b$ has the advantage that 
the cross section can be calculated reliably because the quark and antiquark 
structure functions at the relevant value of $x$ are better known than the 
gluon structure functions that enter the calculation for the W-gluon 
cross section.

In this paper we will concentrate on the single top production process via 
$q\overline{q}'\rightarrow t\overline b$ at the Fermilab Tevatron. In the 
standard model ( SM ), this process can be reliably predicted and the 
theoretical uncertainty in the cross section is only about a few percent 
because of the QCD corrections [9]. Although the statistical error in the 
measured cross section for this process at the Tevatron will be about 
$\pm$ 30\% [8], a high-luminosity would allow a measurement of the cross 
section with a statistical uncertainty of about 1\% [9]. At this level of 
experimental accuracy a calculation of the radiative corrections is 
necessary. In Ref.[9], the QCD and Yukawa corrections to single-top-quark 
production via $q\overline{q}'\rightarrow t\overline b$ have been calculated 
in the SM. But the QCD corrections were found to be quite large, the Yukawa 
corrections were found to be negligible. The SM electroweak corrections are 
also negligible since which are expected to be comparable to the Yukawa 
corrections. The corrections to the single top cross
section in the extensions of the standard model, such as the 
two-Higgs-doublet model ( 2HDM ) [10] and the minimal supersymmetric model 
( MSSM ) [11] have been calculated [12]. In this 
paper we will address the single top production at Tevatron via 
$q\overline{q}'\rightarrow t\overline b$ in the one generation technicolor 
model ( OGTM ) [13, 14]. We will find that the maximum one-loop correction 
to the total cross section for the single-top-production from PGBs in the 
OGTM may reach -2.4\% relative to the tree-level cross section, which may  be 
observable at a high-luminosity Tevatron.

The paper is organized as follows: In section II, we give a brief review of 
the OGTM and then calculate the corrections to the cross section for the 
single top production via $q\overline{q}'\rightarrow t\overline b$ at the 
Fermilab Tevatron. Numerical results and discussions are presented in 
section III.

\begin{flushleft}
\section*{II. One-loop corrections from PGBs }
\end{flushleft}

In detail, we consider the OGTM [13, 14], in which the global flavor symmetry 
will break as follows:
\begin{eqnarray}
SU(8)_L \times SU(8)_R \rightarrow SU(8)_{L+R},
\end{eqnarray}
when the technifermion condensate $< \overline T T >\neq 0$ is formed. 
Consequently 63 ( pseudo ) Goldstone bosons would be produced from this 
breaking. When all other interactions but the technicolor are turned off, 
these 63 Goldstone bosons are exactly massless. Three of them are eaten by
gauge bosons and the others acquire masses 
ranging from a few to above 300 GeV when the gauge interactions are turned 
on.

It was known [15] that only the color-singlet $p$ and color-octet $p_8$ 
contribute to the $W\overline{t} b$ vertex. As for the mass ranges of $p$ and 
$p_8$, the electroweak contribution to their masses is theoretically well 
understood and can be reliably computed ( with some dependence on the 
technicolor model ) [16]:
\begin{eqnarray}
m_p\vert_{EW}=5-14 GeV.
\end{eqnarray}

The ETC interactions also contribute to $m_p$, but this kind of contribution 
is very model dependent [16], so that it is very difficult to make precise 
predictions for the light PGB masses. For the $p_8$, the dominant source of 
its mass derives from the strong color interactions. Under the approximation 
of single-gluon exchange the mass of $p_8$ $m_{p_8}$ was estimated [16] 
according to the relation:
\begin{eqnarray}
\frac{m_{p_8}^2}{m^2(\pi^{\pm})-m^2(\pi^0)}=3[\frac{\Lambda_{TC}}
{\Lambda_{QCD}}]^2
\frac{\alpha_s(\Lambda_{TC})}{\alpha_{em}},
\end{eqnarray}
e.g.
\begin{eqnarray}
m_{p_8}\approx 246\times \sqrt{4/N_{TC}} GeV,
\end{eqnarray}
where the $N_{TC}$ is the number of technicolor.

Because the LEP limit on the Higgs bosons $H^\pm$, 
 $m_{H^\pm} > 50\ \ GeV$ [17], also 
applies to the charged color-singlet $p$, we here assume that $m_p=60 GeV$. 
For the mass of the color-octets $p_8$, 
we will consider the range of 
\begin{eqnarray}
m_{p_8}=200-400 GeV.
\end{eqnarray}

Ellis $et \ \ al.$ [15] estimated the Yukawa couplings to the 
ordinary fermions of the PGBs in the OGTM under some simplifying assumptions. 
The Feynman rules needed in the calculation of the effects of the 
virtual PGBs on the $W\overline{t} b$ vertex at one-loop level come from 
Ref.[15]. One should bear in mind that the technipion decay constant 
$F_{\pi}=250 GeV$ in the first paper of 
Ref.[15] is substituted for $F_{\pi}=123 GeV$ for the 
OGTM. 

Because of the lightness of the b quark when compared with the large top 
quark mass, we will partially neglect the mass of the b quark in the 
calculation for the sake of simplicity. We will use dimensional 
regularization to regulate all the ultraviolet divergences in the virtual 
loop corrections and we adopt the on-mass-shell renormalization scheme.

Fig.1(a) is the tree-level Feynman diagram for single-top-quark production 
via $q\overline q'\rightarrow t\overline b$. The PGBs' corrections of order 
$O(m^2_t/F_{\pi}^2)$ to the process $q\overline q'\rightarrow t\overline b$ 
arise from the Feynman diagrams shown in Fig.1(b, c, d, e).

Including the $O(m_t^2/F^2_{\pi})$ PGBs' corrections, the renormalized 
amplitude for $q\overline{q}'\rightarrow t\overline b$ can be written as
\begin{eqnarray}
M_{ren}=i\frac{g^2}{2}\frac{1}{\hat{s}-m_W^2}\overline u_t (p_3)\Gamma_{\mu}
Lv(p_4)\overline v (p_2)\gamma^{\mu}Lu(p_1),
\end{eqnarray}
where $p_1$ and $p_2$ denote the momentum of the incoming quarks $q$ and 
$\overline{q}'$, while $p_3$ and $p_4$ are the momentum of the outgoing 
$t$ and $\overline b$ quarks, $\hat{s}$ is the center-of-mass energy of the 
subprocess, and we denote the left and right-handed projectors by
\begin{eqnarray}
L,R=\frac{1}{2}(1 \mp \gamma_5),
\end{eqnarray}
and $\Gamma_{\mu}$ is given by
\begin{eqnarray}
\Gamma_{\mu}=-i\frac{g}{\sqrt{2}}[\gamma_{\mu}L(1+\frac{1}{2}
\delta {Z_b^L}+\frac{1}{2}\delta {Z_t^L}+F_L+\frac{1}{4}m_tH_L)],
\end{eqnarray}
where the form factors $F_L$,$H_L$ originate from the vertex corrections 
[ Fig.1(d) ]; $\delta {Z_b^L}$ and $\delta {Z_t^L}$ are the left-handed field 
renormalization constants for the b and t quarks respectively, and 
$\delta {Z_b^L}$, $\delta {Z_t^L}$ have the form of 
\begin{eqnarray}
\delta {Z^L_f}=-\Sigma^L_f(m_f^2)-m_f^2[\Sigma^{'L}_f(k^2)+\Sigma^{'R}_f(k^2)
+2\Sigma^{'s}_f(k^2)]\vert_{k^2=m_f^2},
\end{eqnarray}
where $\Sigma'_f(k^2)=\frac{\partial}{\partial k^2}\Sigma_f(k^2)$, and 
the self-energy ( real part ) has been decomposed according to
\begin{eqnarray}
\Sigma_f(k)=\Sigma^L_f(k)\not k L+ \Sigma^R_f(k)\not k R + m_f\Sigma^s_f(k).
\end{eqnarray}

The renormalized differential cross section of the subprocess is
\begin{eqnarray}
\frac{{\rm d}\hat{\sigma}}{{\rm d}\cos \theta}=\frac{\hat{s}-m_t^2}{32\pi 
{\hat{s}}^2}
\overline {\sum} \vert M_{ren}\vert^2,
\end{eqnarray}
where $\theta$ is the angle between the top quark and incoming quark. 
Integrating this subprocess differential cross section over $\cos \theta$ we 
get 
\begin{eqnarray}
\hat{\sigma}=\hat{\sigma}_0+\delta \hat{\sigma}
\end{eqnarray}
with
\begin{eqnarray}
\hat{\sigma}_0=\frac{g^4}{128\pi}\frac{\hat{s}-m_t^2}{\hat{s}^2(\hat{s}
-m_W^2)^2}[\frac{1}{3}(2\hat{s}^2-m_t\hat{s}-m_t^4)]
\end{eqnarray}
being the tree-level result, and
\begin{eqnarray}
\delta \hat{\sigma}&=&(\delta Z_b^L+\delta Z_t^L+2F_L+\frac{1}{2}m_t H_L)
\hat{\sigma_0}\nonumber\\
&=&\delta \hat{\sigma}_p+\delta \hat{\sigma}_{p_8},
\end{eqnarray}
where $\delta \hat{\sigma}_p$ and $\delta \hat{\sigma}_{p_8}$ stand for the 
contributions of color-singlet PGBs and color-octet PGBs, respectively. The 
explicit forms of $\delta \hat{\sigma}_p$ and $\delta \hat{\sigma}_{p_8}$ are
\begin{eqnarray} 
&&\delta \hat{\sigma}_p=\frac{m_t^2\hat{\sigma}_0}{24F_{\pi}^2\pi^2}
Re[B_1(m_b,m_t,m_p)+B_1(m_t,m_t,m_p)+2m_t^2B_1'(m_t,m_t,m_p)\nonumber \\
&&+m_t^2B_1'(m_t,m_b,m_p)
+2m_t^2B_0'(m_t,m_t,m_p)+4C_{24}+m_t(C_{22}-C_{23}-C_{11}+C_{12})]
\end{eqnarray}
with $C_{ij}=C_{ij}(m_b,m_t,\sqrt{\hat{s}},m_p,m_t,m_p)$ and 
$B_i'=\frac{\partial}
{\partial k^2}B_i$, where functions $B_1$, $B_0$ and $C_{ij}$ 
can be found in Ref.[18], and 
\begin{eqnarray}
\delta \hat{\sigma}_{p_8}=18\delta \hat{\sigma}_p\vert_{m_p \rightarrow 
m_{p_8}}.
\end{eqnarray}
It can be easily found from eqs.(15, 16) that all the ultraviolet divergences 
are canceled for $p$, $p_8$ respectively and therefore the results are 
finite.

The hadronic cross section is obtained by convoluting the subprocess cross 
section $\hat{\sigma_{ij}}$ of partons $i$ and $j$ with parton distribution 
functions $f_i^A(x_1,Q)$ and $f_j^B(x_2,Q)$, which is given by
\begin{eqnarray}
\sigma(s)&=&\sum_{i,j}\int {\rm d}x_1{\rm d}x_2[f_i^A(x_1,Q)f_j^B(x_2,Q)+
(A\leftrightarrow B)]\hat{\sigma}_{ij}(\hat{s},\alpha_s(\mu))\nonumber \\
&=&\sum_{i,j}\int_{\tau_0}^{1}\frac{{\rm d}\tau}{\tau}(\frac{1}{s}
\frac{{\rm d}L_{ij}}{{\rm d}\tau})(\hat{s}\hat{\sigma}_{ij})
\end{eqnarray}
with
\begin{eqnarray}
\frac{{\rm d}L_{ij}}{{\rm d}\tau}=\int_{\tau}^{1}\frac{{\rm d}x_1}{x_1}
[f_i^A(x_1,Q)f_j^B(\tau/x_1,Q)+(A\leftrightarrow B)].
\end{eqnarray}
In the above the sum runs over all incoming partons carrying a fraction of 
the proton and antiproton momenta ( $p_{1,2}=x_{1,2}p_{1,2}$ ), 
$\sqrt{s}=2 TeV$ is the center-of-mass energy of the Tevatron, $\tau=x_1x_2$, 
and $\tau_0=4m_t^2/s$. As in Ref. [19], we do not distinguish the 
factorization scale Q and the renormalization scale $\mu$ and take both as 
$\sqrt{\hat{s}}$. In our numerical calculations, we have used the CTEQ3L 
parton distribution functions [20].

\begin{flushleft}
\section*{III. Numerical results and discussions}
\end{flushleft}

In the following, we present the numerical results for the PGBs' corrections 
to the total cross section for single-top-quark production via 
$q\overline{q}' \rightarrow t\overline b$ at the Fermilab Tevatron with 
$\sqrt{s}=2 TeV$. In our numerical calculations we use $m_Z=91.188GeV$, 
$m_W=80.33 GeV$, $G_F=1.166372\times 10^{-5} GeV^{-2}$, $V_{tb}=1$, 
$\mu=\sqrt{\hat{s}}$, $m_t=176 GeV$, $m_b=4.7 GeV$, $m_p=60 GeV$ and 
$m_{p_8}=200-400 GeV$ as input parameters.

In Fig.2 we plot $\delta\sigma/\sigma_0$ as a function of 
$m_{p_8}$. From Fig.2 one can find that the maximum correction to 
the total cross section may reach -2.4\%, which is expected to be observable 
at a high-luminosity Tevatron.

To summarize, we calculated the one-loop corrections to the single top 
production via $q\overline{q}'\rightarrow t\overline b$ at the 
Fermilab Tevatron from the PGBs in the one-generation technicolor model. 
We found that the maximum correction to the total cross section could 
reach -2.4\%, which may  be observable at a high-luminosity Tevatron.

\begin{flushleft}
\section*{Acknowledgment}
\end{flushleft}

We would like to thank Professor Xinmin Zhang for suggestion of this topic. 
One of the authors Yigang Cao would like to thank Professor Xuelei Wang for 
helpful discussions. This work is supported by the National Natural 
Science Foundation of China and the Natural Science  
Foundation of Henan Scientific Committee.

\newpage
\begin {center}
{\bf Reference}
\end {center}
\begin{enumerate}

\item
F. Abe $et\ \ al.$, The CDF Collaboration, Phys. Rev. Lett. 74 ( 1995 ) 2626; 
S. Abachi $et\ \ al.$, The D0 Collaboration, Phys. Rev. Lett. 74 ( 1995 ) 
2632; S.Willenbrock, HEP-PH/9608418.
\item
R. D. Peccei and X. Zhang, Nucl. Phys. B 337, ( 1990 ) 269; R. D. Peccei, S. 
Peris and X. Zhang, Nucl. Phys. B 349, ( 1991 ) 305; C. T. Hill and S. 
Parke, Phys. Rev. D 49, ( 1994 ) 4454; D. Atwood, A. Kagan and T. Rizzo, 
Phys. Rev. D 52, ( 1995 ) 6264; D. O. Carlson, E. Malkawi and C.-P. Yuan, 
Phys. Lett. B 337, ( 1994 ) 145; S. Dawson and G. Valencia, Phys. Rev. D 53, 
( 1996 ) 1721; T. Han, R. D. Peccei and X. Zhang, Nucl. Phys. B 454, 
( 1995 ) 527; X. Zhang and B.-L. Young, Phys. Rev. D 51, ( 1995 ) 6584; 
G. Gounaris, D. Papadamou and F. Renard, hep-ph/9611224; A. Heinson 
$et \ \  al.$, hep-ph/9612424;
\item
For example, see M. E. Peskin, in Physics and Experiments With Linear 
Collider, Proceedings of the Workshop, Saarilka, Finland, 1991, edited by 
R. Orava and M. Nordberg ( World Scientific, Singapore, 1992 ) P. 1.
\item
F. Berends, J. Tausk and W. Giele, Phys. Rev. D 47 ( 1993 ) 2746.
\item
S. Willenbrock and D. Dicus, Phys. Rev. D 34 ( 1986 ) 155; 
R. K. Ellis and S. Parke, Phys. Rev. D 46 ( 1992 ) 3785; 
G. Bordes and B. Van Eijk, Nucl. Phys. B 435, ( 1995 ) 23.
\item
C. P. Yuan, Phys. Rev. D 41 ( 1990 ) 42. 
\item
S. Cortese and R. Petronzio, Phys. Lett. B 306 ( 1993 ) 386.
\item
T. Stelzer and S. Willenbrock, Phys. Lett. B 357 ( 1995 ) 125.
\item
M. Smith and S. Willenbrock, Phys. Rev. D54(1996) 6696.

\item
J.F. Gunion, H.E. Haber, G.Kane and S.Dawson, 
The Higgs Hunters' Guide ( Addison-Wesley, Reading, MA, 1990 ).
\item
H. E. Haber and G. L. Kane, Phys. Rep. 117 ( 1985 ) 75; 
J. F. Gunion and H. E. Haber, Nucl. Phys. B 272 ( 1986 ) 1.
\item
Chongsheng Li, Robert J. Oakes and Jinmin Yang, Phys. Rev D55 (1997)1672,  
hep-ph/9611455.

\item
S. Weinberg, Phys. Rev. D 13 ( 1976 ) 974; 
D 19 ( 1979 ) 1277; L. Suskind, Phys. Rev. 20 ( 1979 ) 2619.
\item
E. Farhi and L. Suskind, Phys. Rev. D 20 ( 1979 ) 3404; 
S. Dimopoulos, Nucl. Phys. B 168 ( 1980 ) 69; 
S. Dimopoulos $et\ \ al.$, $ibid.$ B 176 ( 1980 ) 449.
\item
J.Ellis, M.K.Jaillard, D.V. Nanopoulos, and P.Sikivie, 
Nucl. Phys. B182 (1981) 529;
E. Eichten $et\ \ al.$, Phys. Rev. D 344 ( 1986 ) 1547; 
S. Dimopoulos, S. Raby and G. L. Kane, Nucl. Phys. B 182 ( 1981 ) 77; 
F. Hayot, Nucl. Phys. B 191 ( 1981 ) 82; 
W. C. Kuo and Bing-lin Young, Phys. Rev. D 42 ( 1990 ) 2274; 
D. Slaven, Bing-lin Young and X. Zhang, $ibid.$ 45 ( 1992 ) 4349.
\item
M. E. Peskin, Nucl. Phys. B 175 ( 1980 ) 197; 
J. Preskill Nucl.Phys. B 177 ( 1981 ) 21; 
P. Binetruy $et\ \ al.$, Phys. Lett. B 107 ( 1981 ) 425.
\item
W. de Boer $\it et al.$, hep-ph/9609209.
\item
M. Clements $et\ \ al.$, Phys. Rev. D 27 ( 1983 ) 570; 
A. Axelrod, Nucl. Phys. B 209, ( 1994 ) 349; 
W. Hollik, Fortschr. Phys. 38 (1990)165.
\item
W. Beenakker $et\ \ al.$, Nucl. Phys. B 411 ( 1994 ) 343.
\item
H. L. Lai $et\ \ al.$, Phys. Rev. D 51 ( 1995 ) 4763.
\end{enumerate}

\newpage

\begin{center}
{\bf Figure captions}                    
\end{center}

Fig.1(a): The Feynman diagram of tree-level $q\overline q'\rightarrow t
\overline b$ process.

Fig.1(b, c, d, e): The Feynman diagrams of one-loop corrections to the 
$W\overline t b$ vertex from PGBs in the OGTM . 

Fig.2: The plot of $\delta\sigma/\sigma_0$ versus $m_{p_8}$ in 
the OGTM ( assuming $m_p=60 GeV$ ).

\newpage

\begin{picture}(30,0)
{\bf 
\setlength{\unitlength}{0.1in}
\put(17,-3){\vector(3,-2){3}}
\put(23,-7){\line(-3,2){6}}
\put(23,-7){\vector(-3,-2){3}}
\put(23,-7){\line(-3,-2){6}}
\multiput(23,-7.5)(0.5,0){26}{$v$}
\put(36,-7){\vector(3,2){3}}
\put(36,-7){\line(3,2){6}}
\put(42,-11){\vector(-3,2){3}}
\put(36,-7){\line(3,-2){6}}
\put(16,-4){$u$}  
\put(16,-10){$\overline{d}$}
\put(43,-4){$t$}
\put(43,-10){$\overline b$}
\put(28,-6){$W^+$}
\put(28,-16){(a)}
\put(5,-35){\line(1,0){18}}
\put(7,-36.5){t}
\put(13,-36.5){t}
\put(19,-36.5){t}
\put(10, -30.5){$p^0$,$p_8^0$,$p^3$,$p_8^3$}  
\put(6,-35){\vector(1,0){2}}
\put(6,-35){\vector(1,0){8}}
\put(6,-35){\vector(1,0){14}}
\put(13,-35){\oval(7,7)[t]}
\put(12,-44){(b)}
\put(33,-35){\line(1,0){18}}
\put(35,-36.5){t}
\put(41,-36.5){b}
\put(47,-36.5){t}
\put(40, -30.5){$p^+$,$p_8^+$}  
\put(34,-35){\vector(1,0){2}}
\put(34,-35){\vector(1,0){8}}
\put(34,-35){\vector(1,0){14}}
\put(41,-35){\oval(7,7)[t]}
\put(40,-44){(c)}
\put(5,-60){\line(1,0){18}}
\put(7,-61.5){b}
\put(13,-61.5){t}
\put(19,-61.5){b}
\put(10, -55.5){$p^+$,$p_8^+$}  
\put(6,-60){\vector(1,0){2}}
\put(6,-60){\vector(1,0){8}}
\put(6,-60){\vector(1,0){14}}
\put(13,-60){\oval(7,7)[t]}
\put(12,-72){(d)}
\multiput(35,-60.5)(0.5,0){14}{$v$}
\put(48,-64){\vector(0,1){4}}
\put(48,-64){\line(0,1){8}}
\put(51,-66){\vector(-3,2){1}}
\put(48,-64){\line(3,-2){4}}
\put(48,-56){\vector(3,2){3}}
\put(48,-56){\line(3,2){4}}
\put(42,-60){\line(3,2){1.4}}
\put(43.5,-59){\line(3,2){1.4}}
\put(45,-58){\line(3,2){1.4}}
\put(46.5,-57){\line(3,2){1.4}}
\put(42,-60){\line(3,-2){1.4}}
\put(43.5,-61){\line(3,-2){1.4}}
\put(45,-62){\line(3,-2){1.4}}
\put(46.5,-63){\line(3,-2){1.4}}
\put(42,-57){$p^3$,$p_8^3$}
\put(41,-64){$p^+$,$p_8^+$}
\put(37,-59){$W^+$}
\put(51,-56){t}
\put(49,-60){t}
\put(51,-65){$\overline b$}
\put(40,-72){(e)}
\put(28,-80){Fig.1}
}
\
\end{picture}
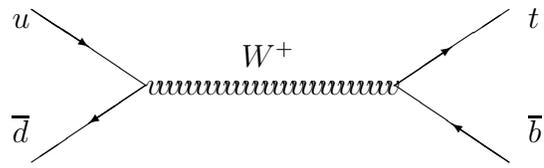
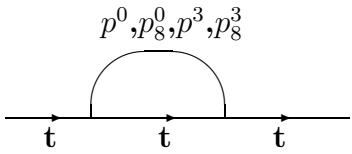
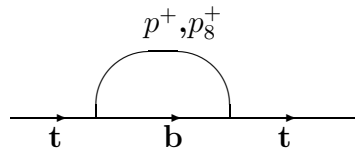
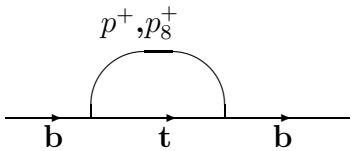
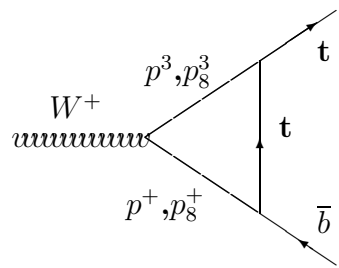
\end{document}